# Engaging Library Users through Social Media Mining

Hongbo Zou[*], Hsuanwei Michelle Chen[§], Sharmistha Dey[¶]
[*]Computer and Information Systems, Queensland University of Technology, Brisbane, AU
[§]School of Information, San José State University, San José, US
[¶]School of Information and Communication Technology, Griffith University, Brisbane, AU
Email: Hongbo.zou@hdr.qut.edu.au; hsuanwei.chen@sjsu.edu; s.dey@griffith.edu.au

**Abstract**

The "participatory library" is an emerging concept which refers to the idea that an integrated library system must allow users to take part in core functions of the library rather than engaging on the periphery. To embrace the participatory idea, libraries have employed many technologies, such as social media to help them build participatory services and engage users. To help librarians understand the impact of emerging technologies on participatory service building, this paper takes social media as an example to explore how to use different engagement strategies that social media provides to engage more users. This paper provides three major contributions to the library system. The libraries can use the resultant engagement strategies to engage its users. Additionally, the best-fit strategy can be inferred and designed based on users' preferences. Lastly, the users' preferences can be understood based on a data analysis of social media. Three such contributions put together to fully address the proposed research question of "how to use different engagement strategies on social media to build participatory library services and better engage more users visiting the library?" [40].

## 1. Introduction

The "participatory library" is an emerging concept, which has come a long way over the past decade. The term refers to the idea that a participatory library, as a truly integrated library system, must allow users to take part in core functions of the library, such as the catalogue system, rather than engaging on the periphery. This concept is been widely adopted in public libraries on their discourse and in practice. *Figure 1* shows the relation of users and libraries with participatory library services. Libraries use new technologies to build their participatory services. And, users visit the built participatory services to communicate with librarian and take part into the management of library. However, how to improve the participatory library services, and better reach the customers to fulfil their needs, is a critical problem faced by libraries. Libraries have employed many emerging technologies, such as web 2.0 and social media, to build their participatory services. Such new technologies can significantly help libraries achieve their mission of engaging with library users, and specifically, allowing users to participate in the conversation with libraries [24]. However, how to use such new technologies to help libraries build their participatory services and engage more users is a new challenge faced by libraries.

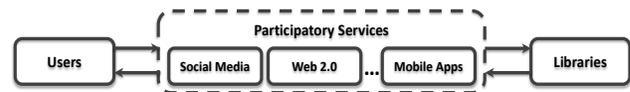

*Figure 1 The relationship between users and libraries with participatory service*

Public libraries notice the changes and are increasingly using social media in an attempt to connect with their users. According to a study published by the Library Research Service (LRS), 93% of U.S. public libraries in all population categories had social media accounts in 2013 [23][29]. Almost all of the large libraries (serving over 500k population), a little more than 4 in 5 middle size libraries (serving the population between 25,000 and 499,999), had at least one social media account in U.S. Libraries make use of social media to engage with their communities [9] or promote libraries services and events [1][25]. Such a trend perfectly matches the concept of the participatory library, which suggests that the library should engage in conversations with its community and inform the community how the library operates [11][34]. With such change trends, social media has become an ideal platform for libraries to create their own participatory services emphasizing user engagement. By having social media channels that are always open and participating in the conversation with users, the library can constantly and effectively evaluate and refine its programs, products, and services to ensure that users are getting what they need [1]. Libraries can take advantage of such social media channels to invite participation, with active rather than passive participation being the goal [16]. Passive participation is when the library provides excellent content and simply asks the user to comment, while active participation involves the library inviting its users to create a community with the library, to help in shaping its direction, co-authoring content and engaging with other users to form a vocal community of users [20].

Despite extensive studies having verified social media potential on enhancing the user's relationship and connection with the library, libraries still need to understand which social media engagement strategies can be used to connect with users, and which of the identified strategies are effective for increasing user engagement, thus for implementing participatory library services. Although social media provides a number of benefits and opportunities for library user engagement, "low-quality" content of social media also affects users' moods and interests negatively. A well-designed strategy should allow the library to better understand users' needs and interests and create more attractive content to engage users. However, how to apply different user engagement strategies to build better channels and engage more users is still being questioned. This study aims to find out what social media engagement strategies are used by public libraries, and which ones are effective for increasing user engagement, thereby implementing participatory library services. This paper intends to explore what social media strategies are used by public libraries in the United States, and which of the identified strategies is effective for increasing user engagement, thus for implementing participatory library services [41][42].

The remainder of this paper is organized as follows. Related work is discussed in Section 2. Section 3 introduces the research approach adopted by this study, and how to design the research to approach the research problem. In section 4, social media mining is involved into phase one study to nail down the research scope and verify the feasibility of research design. In phase two study, section 5 introduces an online survey to verify the findings discovered in phase one study and find out the root causes under facts. The conclusions and limits are discussed in Section 6.

## 2. Related Work

While libraries have traditionally been user focused, the participatory library expands on the radical trust and gives the users more ability to guide the direction of the library service [14]. The public library of the future involves close contact between the library and its users. This participatory library is engaged in conversation with users [26]. By engaging in conversation with users, the library develops knowledge about them that can inform development and delivery of services and collections [11]. This conversational idea also supports the notion of user-driven change, which is often cited as one of the core principles of the future library [1][32]. Social media can support the key ideas that underpin the idea of a participatory library service: user-centered change, participation from users in developing service, and continual re-evaluation of services [1]. Social media also allows the library to enter into the space of the user, rather than waiting for the user to come to it. The library then begins actively seeking out conversations and participation and is able to speak with people it may otherwise not reach [35]. The information and feedback that users provide is the "single best tool" that public libraries have to ensure that they remain relevant [1]. If the key role of the librarian is to "improve society through facilitating knowledge creation in their communities" [11], then librarians must come to understand that a participatory environment is key to facilitating knowledge creation. Social media provides a ready-made communication channel that the library can use to create user engagement and move towards a participatory service [8]. However, there is still a gap between libraries and users through social media communication. The gap is that the topics created by libraries on social media don't necessarily match users' topics of interest in most cases, so the question is: how are libraries using social media to create participatory networks that foster knowledge? [12].

Using social media for library management is an emerging topic, which has gained increased attention in both academia and practice in recent years. Social media can facilitate communications and engagement on library collections and services. Rutherford [24] and Tiffen and England [28] suggest that some libraries are using social media to develop communities and to personalize interactions between the library and users. Tools such as Facebook and Pinterest have been used to build relationships and rapport with client groups [16][20][36], to promote libraries [30], and to provide better information services. The use of social media tools to communicate and to increase engagement can have powerful and positive effects on repeated library visits, rapport-building, referrals or positive feedback[15]. Twitter also provides another new Internet venue to market a library's online brand and impression. There are many libraries that have already created their Twitter communities to connect with their users. While Twitter provides a great avenue for sharing and promotion, it does have its words limits. The library can only post short messages, images or videos, and it is not conducive to detailed discussions. How to understand users solely from responses to social media is a challenging issue and requires an in-depth, innovative data analysis approach.

Although social media provides a good way for libraries connecting with users, a well-studied engagement strategy provides many high-quality participatory services, and can better understand the users' needs. How to plan a good participatory library service? Libraries should consider two things: first, how to use social media to better understand online user behavior, and second, how to use social media to build better library services, thereby engaging more users [34][36]. These two challenges are motivating many of researchers to propose specific research topics to analyze user behavior through social media, and improve participatory services in libraries[18][14]. For example, to figure out the impacts of interactions on social media platform, the information-flow of social media is explored to understand how to use social media to interact with users, and how many interaction types are using social media platform for user engagement[5][18][31]. To understand how libraries perform their engagement strategies, popular social media outlets, such as Pinterest, Twitter, and Facebook, have been studied and examined. The various features and methods of user interactions determine different engagement strategies used [1][22][36]. Also, to explain how do physical library services coexist alongside social media services, some research explores the potential coexistence approaches [1][2][35]. All of referred studies only focus on one or two topics to address the benefit of library social media using. And, they didn't answer the question how to use the social media to improve participatory services and engage more users to participate in library services. All of the related works have been summarized, analyzed, and constructed as a solid base, which support and boost this study to further explore and understand the relationship between social media and library participatory services and answer the question of "how to use different engagement strategies on social media to build participatory library services and better engage more users visiting the library?".

## 3. Research Design

This study aims to explore how libraries understand the needs of users through social media and use social media to improve library participatory services and engage users. The problem includes two research objects: social media and library users. Although social media has become one of the primary ways for the library and users to exchange information, neither the library nor the library user have any well-studied strategy to help them better understand each other. Such a strategy should be a recommendation which can instruct the library using social media to build more attractive participatory services, engage more users, and dictate the user to better understand the library's engagement strategies and participate in library's service more effectively. The deficiency in the study is caused by two challenges. One of the challenges is that social media contains a large amount of unorganized data, which cannot be analyzed manually. The other challenge is that there are no existing engagement strategies to help librarians design a user survey, collect responses and better understand the needs of user. To address these two challenges, the investigation and exploration is conducted on two granularity levels - phase one: social media mining and phase two: a survey study. Figure 2 demonstrates the research design that was developed for use within the study. An overview of the role of each phase in terms of the specific contribution to the overall study is presented in the figure.

Phase one study is designed to use the existing data mining technologies to find out interesting patterns and knowledge. To generalize the study results, 10 public libraries were observed across the United States. These 10 libraries were chosen because of their geographical distribution at different locations and different scales. The data on these 10 public libraries were collected through Twitter in the study. The data set collected from the libraries contains over 10,000 tweets from a time span ranging from several months to several years. Manual analysis cannot be applied on the collected data due to the data scale. Two methods of data mining - topic modelling and sentiment analysis have been introduced and employed to conduct data analysis. Both data mining methods are Natural Language Processing (NLP) tools which can automatically identify topics from a corpus, and label the sentiments from the users' feedback. Four user engagement methods have been extracted by topic modelling technology. Sentiment analysis gives an in-depth analysis on which method is more popular for user engagement. Four such user engagement methods and in-depth analysis become the base of the hypotheses in this study.

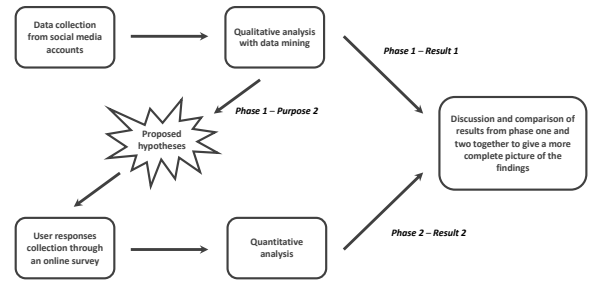

*Figure 2 Research design*

To answer how to use social media to improve participatory services and engage users, research hypotheses are proposed in the phase one. The research hypotheses cover all guesses based on the observation in phase one study, and the explanation of why the guess may be correct. All of the observations are passed to and verified in phase two. Figure 2 shows the connections between phase one and two. To test the hypotheses, an online anonymous survey is designed to collect the feedback from library users in the phase two. Hypotheses are devised to provide a predicted answer to discover how social media improve participatory services with different engagement strategies. The survey includes fifteen questions of which there are thirteen questions related to survey responses and two demographic variables. All of the questions have been designed to generate numerical data. The collected data is quantitatively analyzed with statistical techniques. An in-depth discussion and comparison of findings between phase one and phase two have been conducted at the end of the research. The social media data of 10 geographically distributed libraries have been collected and analyzed in phase one, and similarly, the survey data of phase two has also been collected and labelled with 10 such libraries. Finally, the findings of phase two has verified the proposed hypotheses of phase one.

## 4. Phase One – Social Media Mining

The study aims to investigate what social media engagement strategies are used by public libraries in the United States, and which of the identified strategies is effective for increasing user engagement. To define the existing social media engagement strategies, their scope and to verify the feasibility of research design, the phase one study is conducted with social media mining. The purpose of the phase one study is to classify engagement methods used by public libraries and understand the role of social media in enhancing library participatory services. The phase one study is started with the analysis on the Twitter data of 10 libraries in the United States. A data mining method (topic modelling) is applied to the collected data, and the collected data is classified into four unlabeled categories. To name four such categories, a general study of engagement methods is conducted to define engagement labels. The defined engagement methods were mapped into the four categories classified in topic modelling analysis. Thus, any data analysis and findings in topic modelling could be considered as the consequence of applying different engagement methods. The findings were further checked by a sentiment analysis. Sentiment analysis uses text analysis to determine the attitude of each tweet and which engagement strategy works better on library participatory services.

| Library Name | Location | Join Date | Total tweets | Followers |
|---|---|---|---|---|
| New York Public Library (NYPL) | New York City, NY | Nov. 2008 | 32.5K | 2.6M |
| San José Public Library (SJPL) | San José, CA | Oct. 2009 | 4,475 | 4,480 |
| San Francisco Public Library (SFPL) | San Francisco, CA | Mar. 2009 | 13.1K | 18.4K |
| Los Angeles Public Library (LAPL) | Los Angeles, CA | Apr. 2009 | 28K | 25.3K |
| Birmingham Public Library (BMPL) | Birmingham, AL | Mar. 2008 | 18.1K | 15.3K |
| California State Library (CAPL) | Sacramento, CA | May 2009 | 5,293 | 5,616 |
| Seattle Public Library (SEPL) | Seattle, WA | Jun. 2010 | 28.9K | 22.2K |
| Houston Public Library (HTPL) | Houston, TX | May 2007 | 11.7K | 18.7K |
| Columbus Library (OHPL) | Columbus, OH | Dec. 2008 | 7,495 | 25.3K |
| North Dakota State Library (NDPL) | Bismarck, ND | Oct. 2009 | 3,451 | 2,290 |

*Table 1 Data Collected from 10 Selected Libraries*

**4.1. User Engagement Classification with Topic Modeling**

To observe the user-engagement strategies commonly used by libraries, the twitter data of 10 public libraries were collected. The profiles of 10 public libraries are presented in the *Table 1*. The data has been first collected in 2014 and updated in 2019. These 10 public libraries are selected first with a focus on diversity in geographic location: they are distributed from the East Coast to the West Coast, throughout the United States. In addition, some of the selected libraries have high population density, such as the New York Public Library, while others have low population density, such as the North Dakota State Library. The topic modelling clusters the collected tweets into multiple categories, which have been labelled four user engagement strategies. Four user engagement strategies include *literature exhibits*, *engaging topic*, *community building*, and *library showcasing*. Labels definition is based on the engagement purpose rather than actually posted content. *Literature exhibits* mainly share book covers, historic archives, precious literature, video records, and any documentation available in the library. Literature exhibits serve a traditional library purpose, wherein the library delivers information to its users, and users passively consume those library posts. *Engaging topic* needs the librarians to create appealing topics and content for users. In the meantime, users can actively retweet or comment on their favorite content in response to improve the topic and its communication. *Community building* is a way for libraries to directly interact with users in their community. Libraries are not simply delivering information to patrons; they are also using Twitter to create a virtual "club" and launch special topics for discussion. Any library community topic of concern can, thus, be launched and discussed by users and their library. *Library showcasing* seeks to keep users up to date about what their local library is doing, what new programs have been launched by the library, etc. Libraries can use library showcasing to collect ideas and suggestions from users on their local services and events. The four labels have been mapped into four topics with words checking in topic modelling.

**4.2. Users Sentiment Analysis**

Topic modelling verified the classification of four library user-engagement strategies. This part addresses the question of which strategies work better on user engagement. To answer this question, another data mining technique, sentiment analysis, is conducted. Sentiment analysis uses text analysis to determine the attitude of a text to some topic or the overall contextual polarity of a document. The sentiment analysis is a supervised learning task that consists of assigning a class label to an unclassified tuple according to an already classified instance set, which is used as a training set for the algorithm [7]. The collected dataset is labelled as positive and negative by "responses" number checking. "Responses" is a new attribute and defined as the sum of "favorites" and "retweets". In addition to the training data labelling, a classification model needs to be selected to conduct sentiment prediction. Three representative classification models are evaluated - Naïve Bayes, K-nearest neighbors (KNN), and random forest. The predictive values are calculated by the classifier and compared with the labelled value. Five accuracy features, including true positive rate, false positive rate, precision, f-measure, and receiver operating characteristic (RoC) area are derived by the classification. The detailed accuracy estimation is presented in Table 2. As a result, the random forest classifier is chosen for the sentiment prediction.

| Classifier | Naïve Bayes | 3NN | Random Forest |
|---|---|---|---|
| Accuracy | 82.33% | 81.33% | 84.58% |
| ROC Area | Fair | Fair | Good |

*Table 2 Classifier accuracy comparison*

With the algorithm profiling, the random forest classifier predicts sentiment for each tweet in the unlabeled dataset. Figure 3 shows the sentiment distribution of four engagement strategies for every library. The distribution of each topic is presented with the stacked bar chart. The results show that NYPL and NDPL have posted many engaging topic-related tweets to engage users. Especially, NYPL has over 90% tweets that have been reviewed by users with the positive rating. CAPL and HTPL use many tweets to exhibit their literature. BMPL and SFPL post many community-building-related tweets. A community-building strategy can engage users with popular topic discussions; however, in general, such a strategy does not show a significant contribution to user engagement. The remaining libraries do not show a specific preference for any particular engagement strategy.

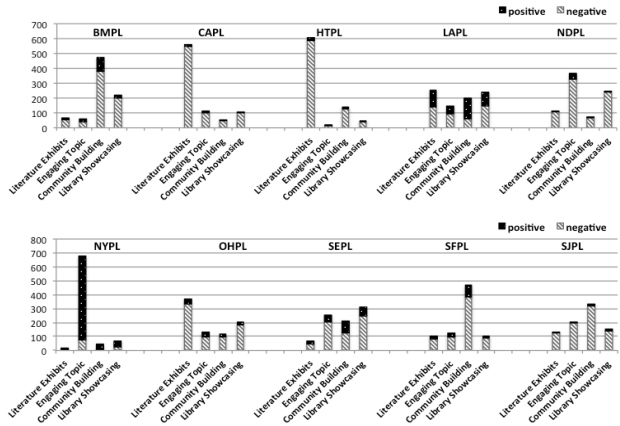

*Figure 3 Sentiment prediction distribution in 10 selected libraries*

### 4.3. Findings in Phase One Study

In the phase one study, the user-engagement methods have been classified into four categories: literature exhibits, engaging topics, community building, and library showcasing. Based on the sentiment analysis of 10 selected libraries, two categories: literature exhibits and engaging topics have been specified by libraries on their participatory services building.

In addition, although social media provides a new platform for user engagement, such engagement needs to be carefully studied and prepared. Even in the literature exhibits, the self-written annotations can greatly help users to pick the books. Libraries should not use Twitter only to extend the traditional mission of the library, in which the library needs to make the collection of sources of information and resource accessible to a defined community for reference or borrowing. The library should creatively launch popular-trends-related topics and share newly created themes for users via Twitter. Favorites, retweets, and comments can help the library better understand the needs of their users and better improve their services to engage the users.

### 4.4. Research Hypotheses

The findings of phase one study expose the existing user engagement strategies used by public libraries, and the impacts of different strategies. Such classification is based on the observation of the data collected from Twitter, which doesn't present the reasons why different strategies have different impacts. So, such classification can moderately help public libraries improve their participatory services. However, this study aims to explore and present the reasons for different impacts. The exploration of the reasons has been guessed in six testable sub-hypotheses. The sub-hypotheses include investigation on how to use four user-engagement strategies to build the social media content, and how to use popular topic and visualization - two methods to organize the format of social media publishing.

Figure 4 presents the conceptual framework of the hypotheses. In the Figure 4, the boxes are the key variables, H's are the hypotheses, and + or – stand for whether it is a positive or negative relationship. Following the above analysis, six types of variables have been presented in the framework. Four engagement strategy variables have the positive perception and could influence user engagement. The rest two variables - topic popularity and visualization describe how libraries organize their social media in content and style, which also have a positive perception. The survey of the study is based on the proposed sub-hypotheses.

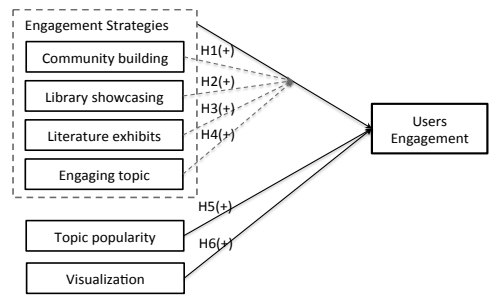

*Figure 4 Conceptual framework of the hypotheses*

The phase one study defines the current landscape and proposes relevant hypotheses. The proposed hypotheses need to be checked and verified in the phase two study.

## 5. Phase Two – A Survey Study

To test the proposed hypotheses, an anonymous survey is designed to identify what could impact the engagement strategy working on library participatory services. Fifteen survey questions are defined and developed in a way that respondents can easily understand and answer. The survey questions include thirteen questions related to survey response, and two questions related to demographic variables. Some survey questions – such as social media usage, are assessed using the frequency of use measure adapted from Kankanhalli et al [10], with the highest value representing frequency of use throughout the month by a user. To measure user engagement, personal involvement is assessed using Zaichkowsky, and Laurent and kapferer measure of involvement using a likert scale where 1 = strongly agree, and 5 = strongly disagree [13][33]. 5-point semantic scale represented users' willingness to be involved in the library's social media. Additionally, some closed questions were designed to offer simple alternatives such as 'Yes' or 'No'. The respondent is restricted to two exclusive options which are easy to answer. Such questions with alternative answers are introduced to complement the personal involvement measurement by forcing the respondent to choose between given alternatives [26]. The demographic variables of age and library's name are included as controls. The library name can help to cross-check the survey findings in the phase one of the study. The survey was distributed by American Library Association (ALA) mailing list. And the responses were collected over six months. A total of 305 respondents participated in the survey. The collected data was analyzed in three stages: descriptive statistics, correlation analysis for survey questions, and in-depth analysis - comparison between the findings and results of phase one and phase two.

### 5.1. Descriptive Statistics

Descriptive statistics provide simple summaries about the observations that have been made. Such summaries may be either quantitative, i.e. summary statistics, or visual, i.e. simple-to-understand graphs. These summaries may either form the basis of the initial description of the data as part of a more extensive statistical analysis, or they may be sufficient in themselves for a particular investigation [17][19].

Descriptive statistics present the basic features of the data and provide simple summaries of the sample and the measures. Some findings have been observed and listed here:

1. Over 73% of respondents physically visit public libraries at least once every month, and only 54% respondents use social media to connect with the library;
2. Over 80% of respondents believe library community building on social media can engage more library users. And only 46% of respondents like to participate in the library community events;
3. Over 82% of respondents believe library showcasing on social media can engage more library users. Only 43% of respondents are interested in participating in the library change discussion on social media;
4. Over 83% of respondents believe literature exhibits can engage more library users. And over 85% of respondents confirm multimedia can help library engage more users;
5. 79.34% of respondents have at least one interested topic they want to learn about from library social media. And only 56% of respondents have been attracted by a discussion topic posted on social media by the library;
6. Only 57% of respondents believe popular topics can engage more library users;
7. 68% of respondents believe visualization is easier to catch the user's attention and engage users.

### 5.2. Diagram Analysis for Survey Questions

Descriptive statistics provide simple summaries about the survey. Diagram analysis is a method of statistical evaluation used to study the strength of a relationship between two, numerically measured, continuous variables. This particular type of analysis is useful when a researcher wants to establish if there are possible connections between variables. In the study, there are total eight variables - including seven variables in the hypotheses and age group. The diagram analysis cannot cover all of the variable combinations which is over 56 combinations. The diagram analysis examines the relationship between variables with aggregations and diagrams, which can filter out some of the examinations of variable combination. The aggregations form the basis of the initial description in the in-depth correlation analysis.

In descriptive statistics, there are over 73% respondents physically visiting the public library at least once every month, and only 54% of respondents using social media to connect with the library. A diagram analysis is needed to check the relation between social media connection and physically visiting. Hopefully, the result of the correlation analysis can answer the questions of whether social media is really effective in improving library participatory services, and which age group is most likely to be impacted by the social media connection. Two conclusions are derived as follows:

1. Social media is really effective in improving library participatory services;
2. Social media engagement has similar impacts on different age groups.

Four engagement strategies have been added into the hypotheses as four variables. Therefore, these four strategies also have been tested in the survey. As for the engaging topic, the survey doesn't ask the question of whether engaging topic on social media can engage more users, because the engaging topic cannot be defined and understood without the specified theme. An alternative question "what topic(s) are you interested in learning about on library social media?" is designed and asked in the survey. There are over 79% respondents who have at least one interested topic on library social media. And also, there are 56% respondents who have been attracted by a discussion topic posted on social media by the library. A diagram checking is needed to verify whether users visiting the library are the same users engaged by social media. Two conclusions are derived as follows:

1. Most people (over 80%) believe the four strategies can engage more users through social media;
2. Some people (over 40%) prefer to connect with the library through social media instead of physically visiting.

In the descriptive statistics, only 176(57%) respondents believe popular topics can engage the users, and 209(68%) respondents believe visualization can easily catch user's attention and engage more users. However, there are 242(over 79%) respondents having at least one interested topic they want to learn from library social media. And also, 260(over 85%) respondents confirm that multimedia content can attract more visitors to the library. To check the relationship of visualization on user engagement, the intersection of two users' groups - 209 respondents believing visualization and 260 respondents confirming multimedia have been checked. Over 198 respondents provided positive answers for both questions. 198 respondents account for 94.73% of all respondents believing in visualization. The findings list is as follows:

1. Respondents believe visualization can engage more users;
2. A popular topic can engage more users, and libraries should create more popular topics users are interested in.

### 5.3. Regression Analysis

With the examination of diagram analysis, there are only eight correlation of variable combinations that need to be analyzed - six hypothesized independent variables and two age related dependent variables (including social media connection and library visit frequency). An in-depth correlation analysis - regression analysis has been selected to examine the relationship between two variables of interests. The regression analysis includes many different regression models for data fitting, such as linear regression, polynomial regression, logistical regression etc. Linear regression has been chosen for the correlation analysis due to the study purpose of linear relationships exploration. Most commonly, linear regression analysis estimates the conditional expectation of the dependent variable given the independent variables - that is, the average value of the dependent variable when the independent variables are fixed. In the fitting, a function of the independent variables, called the regression function, is to be estimated. In regression analysis, it is also of interest to characterize the variation of the dependent variable around the prediction of the regression function using a probability distribution. Some statistical measures have been listed and defined here for the regression analysis.

**Dependent Variable** is the main factor that the analysis is trying to understand or predict.

**Independent Variables** are the factors that the analysis hypothesizes as having an impact on the dependent variable.

**Multiple R** is the Correlation Coefficient that measures the strength of a linear relationship between two variables.

**R Square** is the Coefficient of Determination, which is used as an indicator of the goodness of fit. It shows how many points fall on the regression line.

**Adjusted R Square** is a modified version of R square that has been adjusted for the number of independent variables.

**Standard Error** is the square root of the variance of the regression coefficient. It shows the precision of the regression.

**Observations** is simply the number of observations.

Correlation between social media visits and user engagement is analyzed and presented first. This analysis covers two survey questions. These questions include how often you use social media to connect with the local library last month and the user engagement. The user engagement survey is designed as a series of measurable questions with answers of "strongly disagree", "disagree", "agree", and "strongly agree". With the designed survey questions, such scale stand for the opinion of respondents on social media engagement.

| Regression Statistics | |
|---|---|
| Multiple R | 0.90822912 |
| R Square | 0.82488014 |
| Adjusted R Square | 0.83178981 |
| Standard Error | 0.10041476 |
| Observations | 305 |

*Table 3 The correlation between social media visits and user engagement*

The linear regression analysis results are shown in Table 3. In the analysis, independent variable (X) is defined as social media connection, and the dependent variable (Y) is defined as user engagement. The analysis results are derived from 305 observations. Multiple R is over 0.9, which indicates a perfect positive relationship between X and Y. And here, both R Square and Adjusted R Square are large than 0.8, which means that over 80% of the collected data fit the model. Standard Error shows the precision of the regression analysis, in which the loss ratio is around 0.1. With the analysis results, the clear conclusion is derived that social media can significantly impact user engagement [40].

User age is an important independent variable for user engagement. Especially, social media has a different popularity in the age groups. Diagram analysis also questions the importance of age on user engagement with social media. User engagement with social media includes two measurable dependent variables - social media connection and library visit frequency. This part examines the correlation between age and social media connection. Three questions have covered this analysis in the survey. The questions include how old are you, how many times do you physically visit a public library every month, and how often did you use social media to connect with the local public library last month? These three questions are marked as required in the survey. So, the age of respondents can be connected with their habits on social media connect and library visit frequency.

*Regression Statistics*

| Multiple R | 0.68495427 |
|---|---|
| R Square | 0.53518065 |
| Adjusted R Square | 0.53188032 |
| Standard Error | 0.20047468 |
| Observations | 305 |

*Table 4 The correlation between age and social media visits*

The correlation analyses have been presented in Table 4 and Table 5. Table 4 shows the results of linear regression analysis between age and social media connection. The age is defined as independent variable (X), and social media connection is defined as dependent variable (Y). The analysis results also have been derived from 305 observations. Multiple R is only around 0.68, which indicates a slightly positive relationship between X and Y. And here, both R Square and Adjusted R Square are around 0.5, which means that over 50% of the collected data fit the model. Standard Error shows the precision of the regression analysis, in which the loss ratio is around 0.2. With the analysis results, the clear conclusion is derived that age has minimal impact on social media connection. In other words, all the users prefer to access social media which does not discriminate against age.

*Regression Statistics*

| Multiple R | 0.45998878 |
|---|---|
| R Square | 0.21158967 |
| Adjusted R Square | 0.20828934 |
| Standard Error | 0.16110907 |
| Observations | 305 |

*Table 5 The correlation between age and library visits*

Table 5 shows the results of linear regression analysis between age and library visit frequency. The age is defined as independent variable (X), and library visit frequency is defined as dependent variable (Y). The analysis results are also derived from 305 observations. Multiple R is only around 0.45, which indicates a less positive relationship between X and Y. And here, both R Square and Adjusted R Square are around 0.2, which means that only 20% of the collected data fit the model. Standard Error shows the precision of the regression analysis, in which the loss ratio is around 0.16. With the analysis results, the clear conclusion is derived that age has little impact on library visiting frequency.

|  | Community Building | Library Showcasing | Literature Exhibits | Engaging Topic | Topic Popularity | Visualization |
|---|---|---|---|---|---|---|
| Multiple R | 0.44728069 | 0.33072396 | 0.39216202 | 0.81556986 | 0.81497027 | 0.89772396 |
| R Square | 0.40006001 | 0.30937834 | 0.35379105 | 0.77269831 | 0.76417654 | 0.80590831 |

*Table 6 The correlation analysis between hypothesis variables with user engagement*

Table 6 shows the analysis of linear regression between six hypothesized variables and user engagement. In the analysis, community building, library showcasing, literature exhibits, engaging topic, topic popularity, and visualization have been defined as independent variable (X) respectively. And user engagement is defined as a dependent variable (Y). Every X is examined with the variable Y in turn. The correlation analysis targets to understand which independent variables have significant impact on user engagement. The analysis identifies that three independent variables - engaging topic, topic popularity, and visualization have a significant impact on user engagement. Multiple R of three variables are large than 0.8, which indicates a very positive relationship between the variable X and Y. And also, R Square of three variables are large than 0.7, which means that over 70% of the collected data fit the model. With the analysis, the clear findings discovered are that engaging topic, topic popularity, and visualization have significantly impact on user engagement.

### 5.4. Findings in Phase Two Study

In summary, two questions are explored in phase one. Q1: What engagement strategy is employed by public libraries on social media? Q2: What are the responses of users facing the employed engagement strategy of every library on social media? The answers to these two questions derived a hypothesis, which attempts to explore and answer the question, "how do libraries use social media to improve their participatory service by engaging users more effectively?". To test the hypotheses, an online survey has been conducted in phase two. The responses of the survey are analyzed with three methods: descriptive statistics, correlation analysis, and results comparison with phase one.

A lot of findings were reveals in the analyses, even though some of them are duplicates in the three methods. All of these findings have been listed here after removing duplicates.

1. Social media is really effective in improving library participatory services and engaging more users;
2. Libraries' engagement strategy efforts on social media can increase user participation;
3. User age has little impact on social media engagement strategies;
4. Visualization and popular topics can engage more users. The libraries should create more popular topics in that users are interested on social media.

## 6. Conclusion and Discussion

In the study, the exploration was conducted on two granularity levels - social media mining and a survey study. Phase one collected data from ten public libraries from United States on one of social media platforms - Twitter. Two data mining methods for the qualitative study were applied to the collected data, and interesting patterns and knowledge found out. Four engagement strategies were observed and defined using topic modeling method. The findings were further checked by sentiment analysis. Every tweet was labelled as a positive or negative sentiment to track the impact of every engagement strategy on library users. The findings in phase one addressed two questions: How libraries engage users through social media? And do social media uses have a positive effect on the library user engagement?

Based on the findings of phase one, hypotheses were then proposed to find the best practices for libraries interested in pursuing social media initiatives to effectively engage their users. Phase two designed an online survey to test hypotheses and find out the root causes under facts. 15 questions were designed to formulate hypotheses and detect the underlying reasons for the different engagement strategies. A total of 305 respondents participated in the survey. Three quantitative study methods were applied to the responses. In the analysis, four major findings were discovered to address the question, "what are the practical implications for libraries using social media?".

Based on the findings of two-phase study, three rules were derived to summarize the study.

R1. The library can use one of the engagement strategies to engage its users through social media.

R2. Users' preference can help the library choose the best-fit strategy engaging it users.

R3. The users' preference can be understood based on the analysis of library social media or survey responses.

Three rules were put together to address the research question of "how can libraries use different engagement strategies on social media to build participatory library services and better engage more users visiting the library?". In this study, the social media is a two-way interaction system, where libraries can send the message to online users and have users respond simultaneously. This two-way interaction can be specified as an active approach on which three rules are applied. Consequently, these three rules with two-way interaction can really make a difference to how a library actually use social media platform to generates leads and influences to engage users.

Some possible limitations of this research may include subjectivity, generalization, and underdevelopment of some provisional concepts. Subjectivity of the researcher may have an influence on the research results to a certain extent. Subjectivity may be in the form of personal opinion, or previous experience that prevent the theory from emerging from the data. This may happen during the time of data collection (i.e. some specific survey questions designed by the researcher), or during data analysis (i.e. if the researcher imposes previous knowledge on the data). Being aware of this possibility, this research adheres to the principles and guidelines pertaining to the research method. With the re-examination of the whole study, three potential limitations are listed here:

1. In the phase one study, Twitter has been selected as a social media platform for user engagement strategy study. Therefore, future research should explore whether similar findings can be generalized to different social media platforms;

2. In the phase two study, the survey was distributed through the American Library Associate. Most of the respondents are library professionals, from which the future research should explore more generic audience;

3. In the phase two study, not all survey respondents come from the 10 selected public libraries. The correlation analysis could be better developed with more respondents in the 10 selected public libraries, in the future.

In addition, the generalization of this research may be a little limited on the selected research methods. Due to the nature of qualitative and quantitative research, the number of research participants is modest. With such modest amount dataset, the selected analysis model could overfit the collected data, which definitely leads to a poor generalization on an extended study. Also, the research focuses on a specific library system, i.e., public libraries. This may limit the generalization of the research results to other library systems (i.e., academic, special, and school libraries) and those libraries in the different countries. And also, during the data collection and analysis stage, this research found some interesting concepts and ideas, but they are not supported by data collected in the designed survey. All the survey questions are designed with measurable answers. Most of the time the respondents cannot fully address their responses on the question with measurable answers. Therefore, they are left out of the theory and remain as potential topics for future research.

REFERENCES

[1] Benn, J., & Mcloughlin, D. (2013), Facing our future: social media takeover, coexistence or resistance? The Integration of Social Media and Reference Services, *International Federation of Library Associations and Institutions*, *World Library and Information Congress: 79the IFLA General Conference and Assembly*, Singapore. Retrieved from http://library.ifla.org/129/1/152-benn-en.pdf.

[2] Bruns, T. A., Using social media in libraries: best practices. edited by Charles Harmon and Michael Messina. *Best Practices in Library Services Series. Public Services Quarterly*, ISBN-13: 978-0810887541, 9(4), 313-314.

[3] Cahill, K., User-generated content and its impact on web-based library services, *Oxford: Chandos Publishing (1 edition), Chandos Information Professional Series*, ISBN-13: 978-1843345343, 2009.

[4] Charnigo, L., & Barnett-Ellis P., Checking out facebook.com: the impact of a digital trend on academic libraries. *Information Technology and Libraries*, 26(1), 23-34. DOI: http://dx.doi.org/10.6017/ital.v26i1.3286, 2007.

[5] Chen, D. Y-T., Chu, S. K-W., & Xu, S-Q., How do libraries use social networking sites to interact with users, *Proceedings of the American Society for Information Science and Technology 2012* (*ASIST 2012*), 49(1), 1-10.

[6] Chen, H. M, Zou, H, & Scott, AL., Improving the analysis and retrieval of digital collections: a topic-based visualization model, *Journal of Information Technology Management* Vol. 27, pp. 82-92, 2016.

[7] Czerny, M., Modern methods for sentiment analysis, URL: https://districtdatalabs.silvrback.com/modern-methods-for-sentiment-analysis, 2015.

[8] Fernandez J., A SWOT analysis for Social Media in Libraries, URL: http://pqasb.pqarchiver.com/infotoday/doc/199913917.html Online 35-37, 2009.

[9] Intellectual Freedom Committee, Social media guidelines for public and academic libraries, Approved by the ALA Intellectual Freedom Committee.

[10] Kankanhalli, A., Tan, B. C. Y., & Wei, K. K.. Contributing knowledge to electronic knowledge repositories: An empirical investigation. *MIS Quarterly*, 29(1), 113-143.

[11] Lankes, R. D., The Atlas of New Librarianship. Cambridge, MA: The MIT Press.

[12] Lankes, R. D., Silverstein, J., Nicholson, S., & Marshall, T., Participatory networks: the library as conversation, *Proceedings of the Sixth International Conference on Conceptions of Library and Information Science – "Featuring and Future"*, Information Technology and Libraries, 12(4), 17-33. 2007.

[13] Laurent, G. and Kapferer J-N., Measuring consumer involvement Profiles, *The Journal of Marketing Research*, pp. 41-53, 1985.

[14] Linh, N. C., Helen, P., Edwards, S. L., Towards an Understanding of the Participatory Library, Library Hi Tech, 30(2) 335-346, 2012.

[15] Loudon, L. and Hall, H. From Triviality to Business Tool: The Case of Twitter in Library and Informaiton Services Delivery. Business Information Review, 27(4), 236-241, 2010.

[16] Mack, D., Behler, A. and Rimland, E. Reaching Students with Facebook: Data and Best Practices. Electronic Journal of Academic and Special Librarianship, 8(2), 2007.

[17] Mann, Prem S., Introductory statistics (2$^{nd}$ ed.). Wiley. ISBN 0-471-31009-3, 1995.

[18] Mongeon, P., Using social and topical distance to analyze information sharing on social media, *Proceddings of the Association for Info. Science and Technology (ASIS&T)*, pp 397-403, Feb. 2019.

[19] Nick, Todd G., Descriptive Statistics. *Topics in biostatistics*. Methods in molecular biology. New York: Springer. doi:10.1007/978-1-59745-530-5_3. ISBN 978-1-58829-531-6, 2007.

[20] Phillips, N. K., Academic Library Use of Facebook: Building Relationships with Students. The Journal of Academic Librarianship, 37(6), 512-522, 2011.

[21] Porter, M., & King, D. L., Inviting participation, *Public Libraries Nov/Dec 2007*, 46(6), 34-36, 2007.

[22] Romero, N. L., ROI. Measuring the social media return on investment in a library. *The Bottom Line: Managing Library Finances*, *24*(2), 145-151. doi>http://dx.doi.org/10.1108/08880451111169223.

[23] Rosa, C. D., Cantrell, J., Havens, A., Hawk, J., Jenkins, L., Gauder, B., Limes, R., Cellentani, D., Dalrymple, T., Olszewski, L. Smith, S. & Storey, T., Sharing, privacy and trust in our networked world: a report to the OCLC membership. *Dublin, OH: OCLC Online Computer Library Center*. Retrieved from http://www.oclc.org/content/dam/oclc/reports/pdfs/sharing.pdf, 2007.

[24] Rutherford, L. L. Building participative library services: the impact of social software use in public libraries. *Library Hi Tech*, 26(3), 2008.

[25] Shafawi, S., Hassan, B., User engagement with social media, implication on the library usage: a case of selected public and academic libraries in Malaysia. *Library Philosophy and Practice*, 5-2018.

[26] Siniscalco, M. T. and Auriat, N., Questionnaire Design, *Quantitative research methods in educational planning*, 2005.

[27] Smeaton, K., Davis, K., Using Social Media to Create a Participatory Library Service: an Australian Study. *Library and Information Research*, 54-76, Vol 38, No 117, 2014.

[28] Tiffen, B. and England, A., Engaging with Clients and Personalising Services at UTS Library: Measuring the Value for Libraries and Their Clients. Australian Library Journal, 60(3), 237-247, 2011.

[29] Wancha, M., & Hofschire, L., U.S. public libraries and the user of web technologies, *Library Research Service*. Retrieved from 2014.

[30] Xia, Z. D., Marketing Library Services Through Facebook groups. Library Management, 30(6/7), 469-478, 2009.

[31] Xu, C., Ouyang F., & Chu, H., The academic library meets Web 2.0: applications and implications. *The Journal of Academic Librarianship*, *35*(4), 324-331. doi>10.1016/j.acalib.2009.04.003.

[32] Yardi, S., Romero, D., Schoenebeck, G. and Boyd, D. Detecting Spam in a Twitter Network. First Monday 15(1): 1-4, 2009.

[33] Zaichkowsky, J. L., Measuring the involvement construct, *The Journal of Consumer Research*, 341-352, 1985.

[34] Zou, H., Chen, M, & Dey, S., Understanding library user engagement strategies through large-scale Twitter analysis, *Proceeding of IEEE Big Data Service and Applications (BigDataService 2015)*. 361-370. doi>10.1109/BigDataService.2015.31, 2015.

[35] Zou, H., Chen, M, & Dey, S., Exploring user engagement strategies and their impacts with social media mining: the case of public libraries, *Journal of Management Analytics*, Vol. 2, Issue 4, pp. 295-313, Dec. 2015. doi>10.1080/23270012.2015.1100969, 2015

[36] Zou, H., Chen, M, & Dey, S., A quantitative analysis of Pinterest: understanding library user engagement strategies for effective social media use, Journal of Information Technology Management, Vol. 26, Issue 3, pp. 21-32, Oct. 2015.

[37] Zou, H., Yu, Y. Tang, W. & Chen, H., Flexanalytics: a flexible data analytics framework for big data applications with I/O performance improvement, Journal of Big Data Research, pp. 4-13, Aug. 2014.

[38] Zou, H., Schwan, K., Slawinska, M., Eisenhauer, G., Zheng, F., Liu, Q., Klasky, S. et. al., FlexQuery: An online query system for interactive remote visual data exploration at large scale, Proceeding of IEEE international conference on Cluster Computing (CLUSTER'13), 2013.

[39] Zou, H., Zheng, F., Eisenhauer, G., et. al., Quality-Aware data management for large scale scientific applications, *Proceeding of ACM/IEEE High Performance Computing, Networking, Storage and Analysis (SC 2012)*, Nov. 2012.

[40] Zou, H., Understanding the role of social media in enhancing partipatory services in public libraries, *Ph.D thesis*, 2019.

[41] Zou, H. etc., A Source-aware Interrupt Scheduling for Modern Parallel I/O Systems, *Processding of ACM/IEEE 26$^{th}$ Parallel and Distributed Processing Symposium* (*IPDPS 2012*), pp.156-166, 2012.

[42] Zheng, F., Zou, H., FlexIO: IO middleware for location-flexible scientific data analytics, *Processding of ACM/IEEE 27$^{th}$ Parallel and Distributed Processing Symposium* (*IPDPS 2013*), pp.156-166, 2013.